\documentclass[twocolumn,preprintnumbers,amsmath,amssymb,prl]{revtex4-1}
\usepackage[dvipdfmx]{graphicx}

\usepackage{graphicx}
\usepackage{dcolumn}
\usepackage{bm}
\usepackage{color}
\usepackage{rotating}
\usepackage{ulem} 

\begin{document}

\title{Quantized and unquantized thermal Hall conductance of  Kitaev spin-liquid candidate $\alpha$-RuCl$_3$}

\author{Y. Kasahara$^1$}
\author{S. Suetsugu$^1$}
\author{T. Asaba$^1$}
\author{S. Kasahara$^{2}$}
\author{T. Shibauchi$^3$}
\author{N. Kurita$^4$}
\author{H. Tanaka$^4$}
\author{Y. Matsuda$^1$}

\affiliation{Department of Physics, Kyoto University, Kyoto 606-8502, Japan}
\affiliation{Research Institute for Interdisciplinary Science, Okayama University, Okayama 700-8530, Japan}
\affiliation{Department of Advanced Materials Science, University of Tokyo, Chiba 277-8561, Japan}
\affiliation{Department of Physics, Tokyo Institute of Technology, Tokyo 152-8551, Japan}



\begin{abstract}
Despite extensive investigations, a topological state that hosts Majorana edge modes in the magnetic field-induced quantum disordered state of the Kitaev candidate material $\alpha$-RuCl$_3$ has been hotly debated.   To gain more insight into this issue, we measured the thermal Hall conductivity $\kappa_{xy}$ of various samples grown by the Bridgman method.     The results show that the half-integer quantum thermal Hall effect is intimately related to the magnitude of longitudinal thermal conductivity and the N\'{e}el temperature at zero field, both of which are sample dependent. Samples exhibiting the half-integer quantum thermal Hall effect have larger zero-field thermal conductivity values than a threshold value, implying that a long mean free path of heat carriers is an important prerequisite. In addition, we find that samples with a higher N\'{e}el temperature exhibit a higher magnetic field at which quantization starts to occur.  These results indicate that the quantization phenomenon is significantly affected by the impurity scatterings and the non-Kitaev interactions. 
\end{abstract}
\maketitle

A quantum spin liquid (QSL) is an exotic state of matter in which spins are quantum mechanically entangled over long distances without symmetry-breaking magnetic order \cite{Balents10}.  The exactly solvable Kitaev model on the honeycomb lattice has recently received enormous interest in the hope of achieving novel QSL states,  where quantum spins are fractionalized into itinerant Majorana fermions and localized $Z_2$ fluxes (visons) \cite{Kitaev06}.  Recently, $\alpha$-RuCl$_3$ is drawing much attention as a promising candidate material hosting a Kitaev QSL \cite{Takagi19}. The structure of  $\alpha$-RuCl$_3$ consists of edge-sharing RuCl$_6$ octahedra that are stacked along the $c$ axis via a van der Waals interaction. This compound is a spin-orbit assisted Mott insulator, in which Ru$^{3+}$ ions have an effective spin-1/2 state forming two-dimensional (2D) honeycomb layers \cite{Plumb14}.  The spin-orbit coupling and edge-sharing octahedra structure allow a bond-dependent Ising interaction between local moments of Ru$^{3+}$ ions \cite{Jackeli09}. Although $\alpha$-RuCl$_3$ exhibits an antiferromagnetic (AFM) ordering with a zigzag spin structure below the N\'{e}el temperature $T_N\sim7.5$\,K due to non-Kitaev interactions, such as Heisenberg exchange and off-diagonal interactions, the Kitaev interaction is predominant among the magnetic interactions \cite{Motome20,Winter17}.  The AFM order vanishes by magnetic fields applied parallel to the $ab$ plane, leading to the appearance of a field-induced quantum disordered (FIQD) state above $\sim 8$\,T.

The fingerprint for Majorana fermions of the fractionalized spin excitations in $\alpha$-RuCl$_3$ has been reported by several experiments.  The specific heat measurements have revealed that the magnitude of the entropy release is quantitatively consistent with the spin fractionalization \cite{Do17,Widmann19}.  The broad magnetic continuum observed by Raman scattering has been interpreted by the presence of fermionic excitations,  in contrast to conventional bosonic magnetic excitations \cite{Sandilands15}. Inelastic neutron scattering (INS) measurements also reported a magnetic continuum \cite{Do17,Banerjee16,Banerjee18,Balz19,Ran22}, which corresponds to the Kitaev interaction.  These results reflect that $\alpha$-RuCl$_3$ locates in the vicinity of the phase described by the Kitaev model.     The Kitaev QSL has topologically protected edge states which are characterized by the Chern number $C_h=\pm1$ of the Majorana bands.  The Chern number is given by the sign of the product $h_xh_yh_z$, where $h_x$, $h_y$, and $h_z$ are the $x$, $y$, and $z$ components of the applied magnetic field ${\bm H}$ along the {\it spin} (not crystallographic) axes, respectively. Recently, when magnetic fields are applied tilted from the $c$ axis, a half-integer quantized thermal Hall (HIQTH) effect has been reported by several groups \cite{Kasahara18,Yokoi21,Yamashita20,Bruin21}; the 2D thermal Hall conductance per honeycomb plane $\kappa_{xy}^{\rm 2D}$ shows a quantized plateau behavior as a function of $H$ and has a half value of $K_0$, where $K_0=\frac{\pi^2k_B^2}{3h}T$ is the quantum thermal conductance.   The HIQTH conductance provides direct evidence of a non-Abelian phase and topologically protected chiral edge modes of charge-neutral Majorana fermions.  Moreover, it has been reported that the field-angular variation of the HIQTH conductance has the same sign structure as $C_h$,  $\kappa_{xy}^{\rm 2D}=\frac{1}{2}C_hK_0$ \cite{Yokoi21}.   In particular,  for {\boldmath $H$}$\parallel a$, the planar thermal Hall effect, which is the finite thermal Hall effect for a magnetic field with no out-of-plane components, appears and exhibits a half-quantized plateau.   In addition to thermal Hall measurements, recent detailed angle-dependent specific heat measurements have revealed the appearance of an excitation gap $\Delta_M$ that increases in proportion to the approximate cube of the magnetic field for {\boldmath $H$}$\parallel a$,  which is consistent with the field dependence of the Majorana gap \cite{Tanaka22}.  In contrast,  for {\boldmath $H$}$\parallel b$,  no thermal Hall effect is observed and specific heat measurements report the emergence of quasiparticles that have a gapless linear dispersion \cite{Yokoi21,Tanaka22}.   These results suggest that the non-Abelian topological order persists even in the presence of non-Kitaev interactions in $\alpha$-RuCl$_3$.   

However, despite intensive research efforts,  the topological properties in $\alpha$-RuCl$_3$, in particular Majorana excitations, are still under debate, with conflicting results from theoretical and experimental studies.   It has been proposed that the magnetic continuum reported by INS can be interpreted by the anharmonic magnons caused by strong magnon-magnon interactions \cite{Winter17NC,Maksimov20}.   The HIQTH conductance has been reproduced in some groups \cite{Yamashita20,Bruin21}, but not in others \cite{Czajka21,Czajka22,Lefrancois21}.  Remarkably, instead of the quantization of $\kappa_{xy}^{\rm 2D}$,  quantum oscillations in the longitudinal thermal conductivity $\kappa_{xx}$ has been reported both in zigzag AFM and FIQD states \cite{Czajka21}, suggesting further exotic quantum spin states.   In addition, it has been shown that the HIQTH conductance disappears at high fields \cite{Kasahara18,Yokoi21}.  This implies that there is a distinct intermediate phase separated from the higher field phase by a topological phase transition within the FIQD state. However, some groups have reported the presence of an intermediate phase \cite{Balz19,Kasahara18,Yokoi21,Tanaka22,Suetsugu22}, while others have claimed the absence of such a phase \cite{Bachus21,Modic21,Ponomaryov20}. 

To shed more light on the above conflicting issues in $\alpha$-RuCl$_3$, it is crucially important to understand in which samples the HIQTH effect is observed and in which samples it is not.  Here, we measured $\kappa_{xy}$ of various samples.    Because it has been shown that different growth methods produce slightly different crystal structures of $\alpha$-RuCl$_3$ 
\cite{Bruin22}, 
we used one type of crystals, all of which are grown by the same method.   We find that the HIQTH  effect is observed only in the samples with large longitudinal thermal conductivity in zero field. We also report that the field range at which the quantization occurs is significantly influenced by the N\'{e}el temperature that is slightly sample-dependent.   We discuss these results in terms of impurity scatterings and non-Kitaev interactions.

\begin{figure}[t]
 	\begin{center}
 		\includegraphics[width=1.0\linewidth]{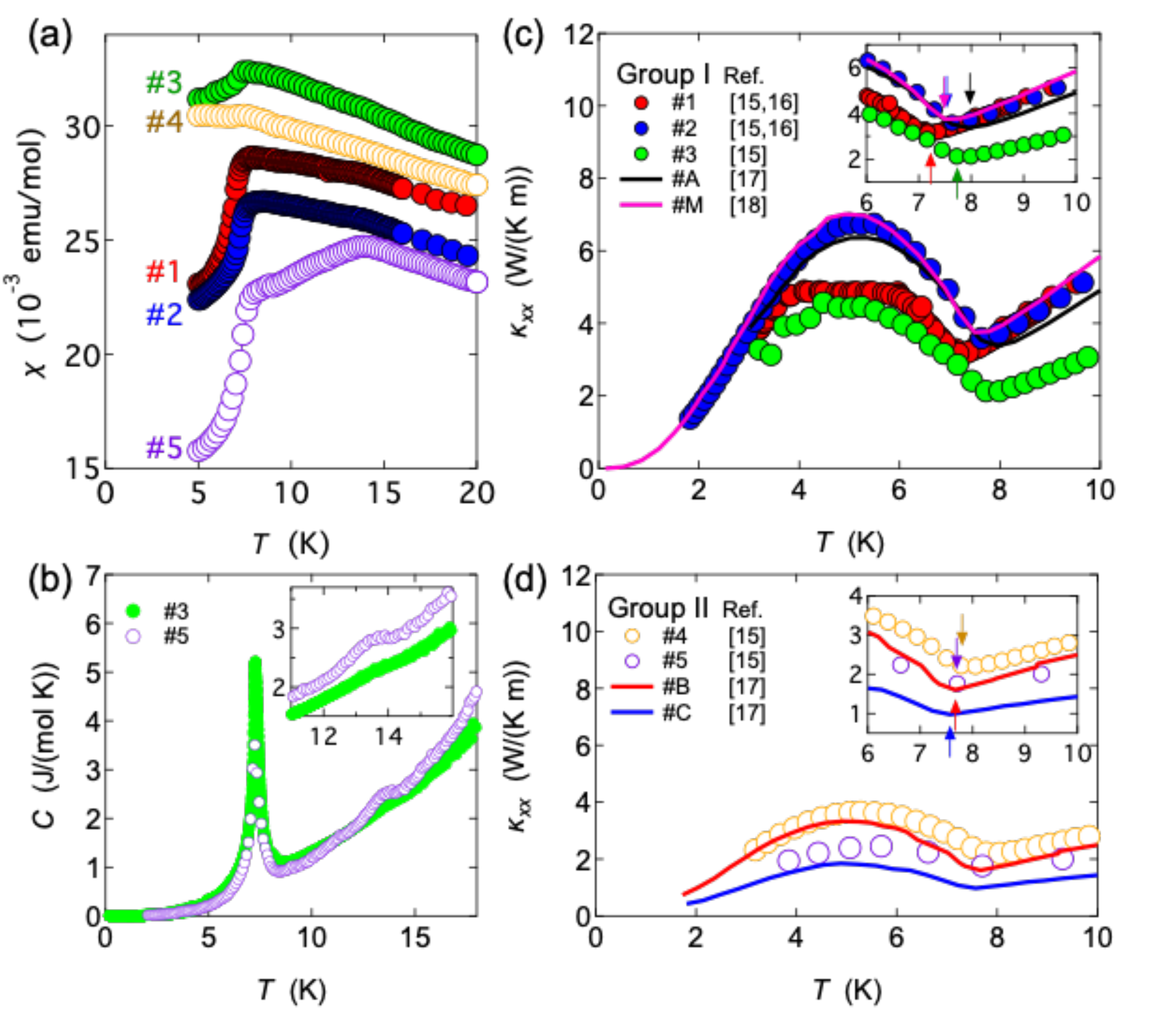}
		\caption{(a) Temperature dependence of magnetic susceptibility $\chi$ under an in-plane magnetic field of 1000\,Oe. 
The data of samples No.\,3 and No.\,4 are vertically shifted for clarity. All the data are the same data shown in Ref.\,\cite{Yokoi21} ($\chi$ of samples No.\,3, No.\,4, and No.\,5 are calculated from the magnetization). 
(b) Specific heat of samples No.\,3 and No.\,5 in zero field. 
The inset shows the data around 14\,K. The data of sample No.\,5 are vertically shifted for clarity. 
(c), (d) Temperature dependence of $\kappa_{xx}$. The data are the same as those reported in Ref.\,\cite{Yokoi21}. The solid lines represent the results reported in Refs.\,\cite{Yamashita20} and \cite{Bruin21}.  The insets show $\kappa_{xx}$ in the vicinity of the N\'{e}el temperature, which are shown by arrows. }
 	\end{center}
 \end{figure}

\begin{figure*}
	\begin{center}
		\includegraphics[width=1.0\linewidth]{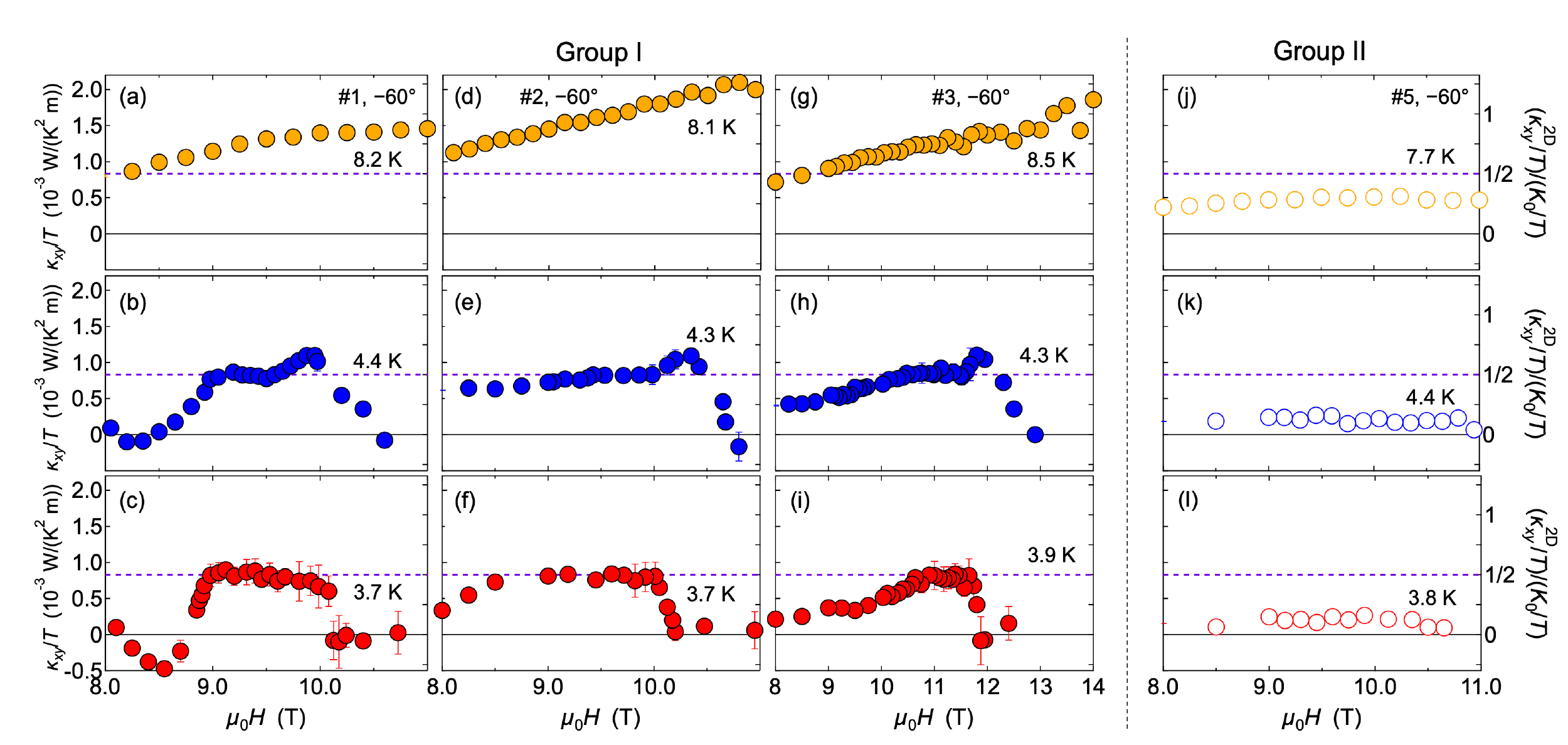}
		\caption{(a)-(c),(d)-(f), (g)-(i) and (j)-(l) show the field dependence of $\kappa_{xy}/T$ at $\theta=-60^\circ$ at different temperatures for samples No.\,1, No.\,2, No.\,3, and No.\,5, respectively. 
Some of the data [(a)-(c), (e), (h), and (k)] are the same data reported in Refs.\,\cite{Yokoi21} and \cite{Kasahara18}. 
The dashed violet lines represent the HIQTH conductance, $\kappa_{xy}^{2D}/T=(K_0/T)/2$.   }
	\end{center}
\end{figure*}

Single crystals of $\alpha$-RuCl$_3$ were grown by the vertical Bridgman method \cite{Kubota15}.   We measured the crystals taken from the same growth batch as those used in the previous reports \cite{Kasahara18,Yokoi21}.  The magnetic susceptibility $\chi$ was measured using a superconducting quantum interference device (SQUID) magnetometer and the specific heat $C$ was measured by a long-relaxation time method \cite{Taylor07}.  We measured the in-plane thermal conductivity $\kappa_{xx}$ and thermal Hall conductivity $\kappa_{xy}$ for five crystals using a setup described in Ref.\,\cite{Yokoi21}.  A heat current ${\bm q}$ was applied along the $a$ axis.  For the measurements of $\kappa_{xy}$, magnetic fields {\boldmath $H$} are applied within the $ac$ plane.

In $\alpha$-RuCl$_3$, the $ABCABC$ stacking arrangement of the 2D layers along the $c$ axis is expected in the trigonal structure, but stacking faults are formed easily and some regions of the sample can crystallize in alternative stacking structures, such as $ABAB$ \cite{Cao16}.  It has been shown that $ABAB$-type stacking faults result in additional local magnetic ordering at $T_{N2}\sim$14\,K \cite{Kubota15,Cao16}.  Figures\,1(a) and 1(b) show the temperature dependence of magnetic susceptibility for samples No.\,1-No.\,5 and specific heat for samples No.\,3 and No.\,5, respectively.    As shown in Fig.\,1(a), $\chi(T)$ in all crystals shows kink transitions at $T_N$.   While no distinct anomaly is observed at $T_{N2}$ for samples No.\,1-No.\,4,  sample No.\,5 exhibits a clear kink anomaly at $T_{N2}$.  The presence of an anomaly at $T_{N2}$ in sample No.\,5 is also confirmed by the specific heat measurements at zero field, while no anomaly is observed in sample No.\,3, as shown in Fig.\,1(b).

Figures\,1(c) and 1(d) depict the temperature dependence of $\kappa_{xx}$ in zero field for samples No.\,1, No.\,2, and No.\,3 and for samples No.\,4 and No.\,5,  respectively.   We note that the magnitude of $\kappa_{xx}$ varies greatly from sample to sample.   Here, we divide the samples into two groups according to the magnitude of $\kappa_{xx}$, the large ones being group I shown in Fig.\,1(c) and the small ones being group II shown in Fig.\,1(d).    In these figures, the results reported in Ref.\,\cite{Yamashita20} (samples A, B, and C) and \cite{Bruin21} (sample M) are also plotted.   For all the samples, $\kappa_{xx}$ shows a sharp kink at $T_N$. Such a sharp kink at the transition has been reported in heavy fermion compounds \cite{Izawa2001,Behnia2005}. Below $T_N$, $\kappa_{xx}$ increases rapidly with decreasing temperature, peaks at around 5\,K, and decreases at lower temperatures.     In this temperature range, the phonon contribution is important in the thermal conductivity.  Upon entering the AFM state below $T_N$, the formation of the magnon gap strongly enhances the phonon mean free path through the suppression of magnon-phonon scattering, giving rise to the enhancement of $\kappa_{xx}$ \cite{Yamashita20,Hentrich18}. Thus the maximum thermal conductivity around 5\,K, which is denoted as $\kappa_{xx}^p$, measures the mean free path of heat carriers. 
We point out that the magnitude of $\kappa_{xx}^p$ is comparable between samples No.\,4 and No.\,5 regardless of the presence or absence of the stacking faults. This indicates that the absence of the stacking faults does not guarantee the high quality of the samples. 
Moreover, although it has been proposed that a sample with larger $\kappa_{xx}$ shows a larger decrease of $\chi(T)$ below $T_N$ \cite{Yamashita20}, such a correlation is not observed in the present experiments [see, for example, $\chi(T)$ and $\kappa_{xx}$ data of samples No.\,1 and No.\,2].
We also note that the  N\'{e}el temperatures of each sample can be sensitively detected by the distinct kink of the thermal conductivity, as indicated by arrows in the insets of Figs.\,1(c) and 1(d).  
To determine $T_N$ accurately, we measured $\kappa_{xx}$ with a temperature gradient $\Delta T$ of 1\,\% or less of the measured temperature ($\Delta T/T<0.01$), especially in the vicinity of $T_N$. We find that the N\'{e}el temperature is slightly sample dependent.  Table\,I lists $T_N$ determined by the thermal conductivity and $\kappa_{xx}^p$ of samples No.\,1-No.\,5, along with those reported in Refs.\,\cite{Yamashita20, Bruin21}. 

\begin{table}[b]
	\caption{The N\'{e}el temperature $T_N$ determined by the thermal conductivity, 
	maximum $\kappa_{xx}$ below $T_N$ in zero field, $\kappa_{xx}^p$ 
	($p$ denotes the peak in $\kappa_{xx}$), 
	angle $\theta$ between applied magnetic field and the $c$ axis, and the field range where the plateau of HIQTH conductance is observed in each samples. The results reported in Refs.\,\cite{Yamashita20,Bruin21} 
	are also shown. }
	\begin{center}
		\scalebox{0.8}[0.9]{
			\begin{tabular}{cccccc}
				\hline\hline
				Sample\ & $T_N$\,(K)\  & $\kappa_{xx}^p$ (W/Km)\ & $\theta$ & plateau field range (T)\ \\
				\hline
				Group I &  & & &  \\
				(quantized) &  & & &  \\
				No.\,1  & 7.2 & 4.9 & $-60^\circ$& 9.0-10.1 \\
				No.\,1  &   &   & $-45^\circ$& 9.4-10.4 \\
				No.\,2 & 7.6  & 6.7 & $-60^\circ$& 9.4-10.0 \\
				No.\,3  & 7.7 & 4.5 & $-60^\circ$& 10.5-11.6 \\
				No.\,3 &  &   &  $+60^\circ$&10.5-13.2 \\
				No.\,3 &   &  &  $-90^\circ$& 9.9-11.3 \\
				A \cite{Yamashita20} & 7.9  & 6.4 & $+45^\circ$ & 12.1-12.7 \\
				M \cite{Bruin21} & 7.5  & 7.0 & $-70^\circ$& \\
				  &  & & &  \\
				Group II &  & & &  \\
				(unquantized) &  & & &  \\
				No.\,4 & 7.8  &  3.7 &$-60^\circ$ &  \\
				No.\,5 & 7.7  & 2.5  & $-60^\circ$&  \\
				No.\,B \cite{Yamashita20} & 7.7  & 3.3 & $+45^\circ$&  \\
				No.\,C \cite{Yamashita20} & 7.6 & 1.8 & $+45^\circ$ &  \\
				\hline
			\end{tabular}
		}
	\end{center}
\end{table}

Figures\,2(a)-2(c), 2(d)-2(f), 2(g)-2(i), and 2(j)-2(l) display the magnetic field dependence of $\kappa_{xy}/T$ for $\theta=-60^\circ$ at several temperatures for samples No.\,1, No.\,2, No.\,3, and No.\,5, respectively, where $\theta$ is the angle between {\boldmath $H$} and the crystallographic $c$ axis. The dashed violet lines represent the HIQTH conductance, $\kappa_{xy}^{2D}/T=\frac{1}{2}C_h(K_0/T)$ with $C_h=+1$.   For samples No.\,1, No.\,2,  and No.\,3,  $\kappa_{xy}/T$ increases with $H$ and is higher than the half quantized value above 8\,K.   At lower temperatures, $\kappa_{xy}/T$ exhibits plateau behavior as a function of $H$ in some field ranges and its value is very close to the HIQTH  conductance.    The plateau field range of  sample No.\,2 is similar to sample No.\,1 but is slightly lower than that of sample No.\,3.  A possible origin of this difference will be discussed later.   In contrast to samples No.\,1, No.\,2, and No.\,3, $\kappa_{xy}/T$ of sample No.\,5 is smaller than the half-quantized value in the whole temperature range.  Thus, the samples belonging to group I exhibit HIQTH  conductance, while those in group II do not show the quantization.

\begin{figure}[t]
	\begin{center}
		\includegraphics[width=1.0\linewidth]{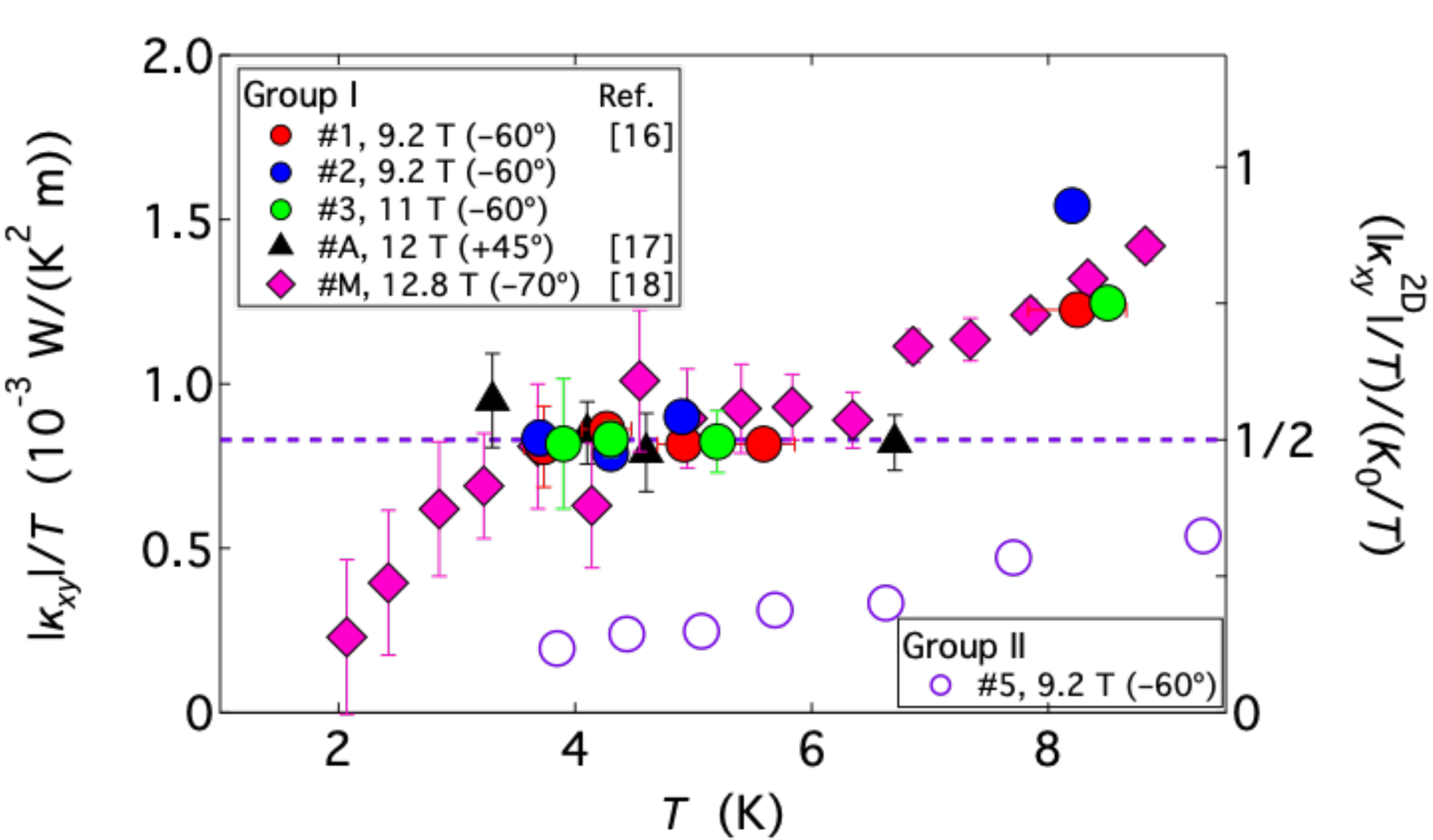}
		\caption{
		Temperature dependence of $|\kappa_{xy}|/T$  for six different $\alpha$-RuCl$_3$ crystals.  The dashed violet lines represent the HIQTH  conductance, $\kappa_{xy}^{2D}/T=(K_0/T)/2$. 
		}
	\end{center}
\end{figure}

\begin{figure}[t]
	\begin{center}
		\includegraphics[width=1.0\linewidth]{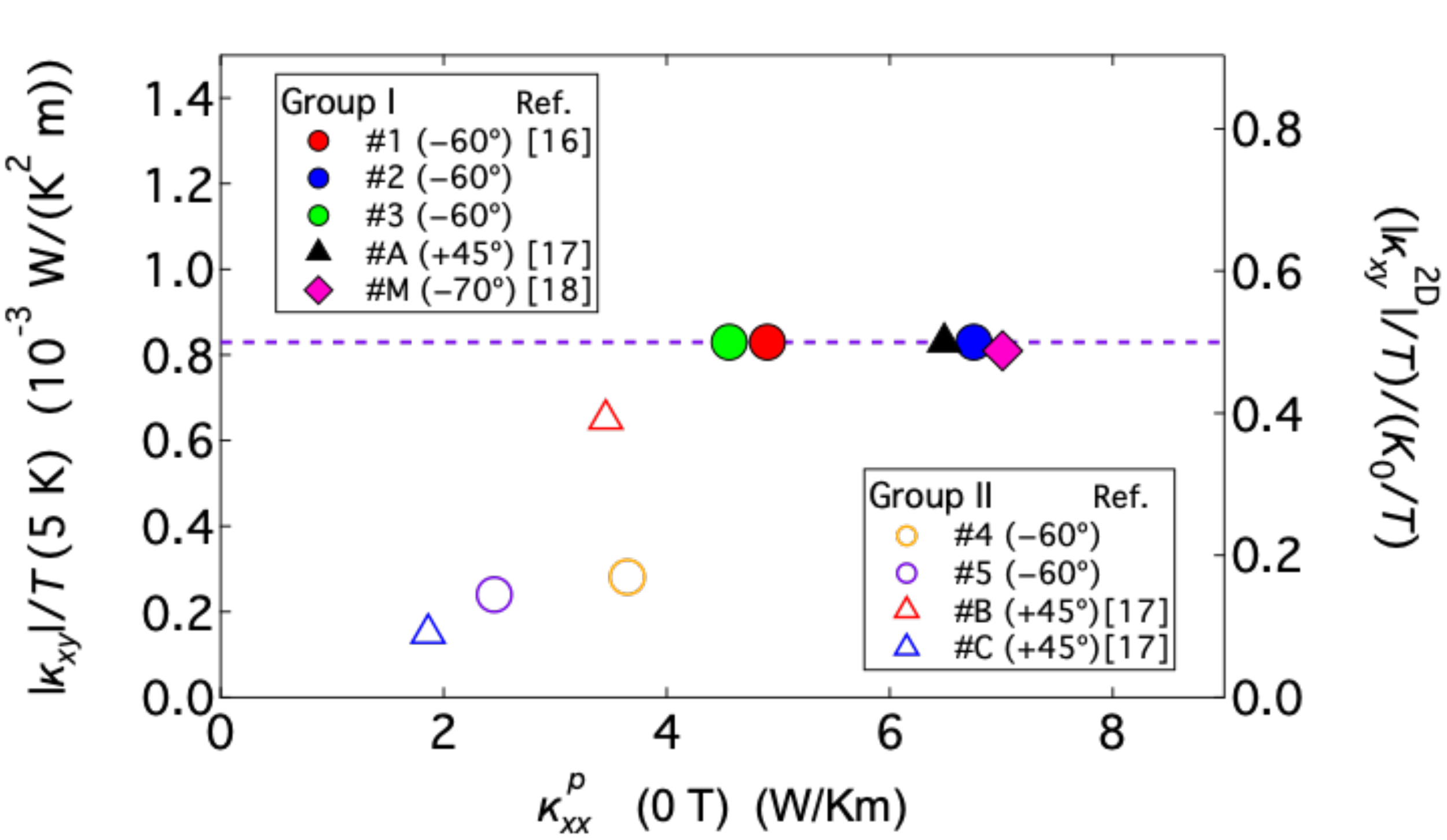}
		\caption{$|\kappa_{xy}|/T$ in the plateau region plotted as a function of 
		maximum $\kappa_{xx}$ below $T_N$ in zero field, $\kappa_{xx}^p$ 
		($p$ denotes the peak in $\kappa_{xx}$). 
		}
	\end{center}
\end{figure}

Figure\,3 shows the temperature dependence of $|\kappa_{xy}|/T$ at the fields where the HIQTH  effect is observed.  Results reported in Refs.\,\cite{Yamashita20,Bruin21} 
are also shown for comparison (samples A and M).   For all the samples that belong to group I, $|\kappa_{xy}|/T$  shows a similar temperature dependence.  As the temperature decreases from high temperatures, $|\kappa_{xy}|/T$ decreases.  Half-integer quantization is observed around 6.5\,K and continues down to around 3.5\,K. At even lower temperatures, it has been reported that $|\kappa_{xy}|/T$ is reduced to be smaller than the half-integer quantization value, as shown by the data for sample M.  In contrast, for samples that belong to group II, 
the $T$ dependence of $|\kappa_{xy}|/T$ is essentially different from that of group I; 
$|\kappa_{xy}|/T$  decreases monotonically with decreasing $T$, and its value is smaller than the half-quantized value in the whole temperature range.    We note that a similar $T$ dependence has been reported in the samples grown by the chemical vapor transport method 
with similar $\kappa_{xx}^p$ values to those of group II 
and no quantization has been observed \cite{Czajka21,Czajka22}. 
This suggests that the sample quality plays an important role in the HIQTH effect. 
However, since the preparation method of the samples is different, a simple comparison should be made with caution. The enhancement of $|\kappa_{xy}|/T$ from the half-quantized value at high temperatures has been attributed to excitations of visons and different types of Majorana fermions in addition to Majorana edge currents \cite{Go19}.  The reduction from the half-quantized value at low temperatures has been discussed in terms of the decoupling phenomena between phonons and Majorana edge currents \cite{Rosch18,Ye18}.

In Fig.\,4, we plot $|\kappa_{xy}|/T$ at 5\,K, where the HIQTH effect is observed, as a function of $\kappa_{xx}^p$. 
Although it has been qualitatively suggested that samples with higher $\kappa_{xx}$ tend to exhibit the HIQTH effect \cite{Yokoi21,Yamashita20}, the present results obtained using the samples from the same batch demonstrate that there is a threshold value $\kappa_{xx}^p\sim4$\,W/Km above which the quantization occurs. 
The presence of the threshold value of $\kappa_{xx}^p$ is clearly seen by comparing samples No.\,3 and No.\,4; sample No.\,3 with $\kappa_{xx}^p=4.5$\,W/Km exhibits the HIQTH effect, while sample No.\,4 with $\kappa_{xx}^p=3.7$\,W/Km does not. 
For samples with smaller values (group II), no quantization is observed, and $|\kappa_{xy}|/T$ at 5\,K decreases with decreasing $\kappa_{xx}^p$. This suggests 
that 
low scattering by impurities and defects is intrinsically important to yield the quantization of the thermal Hall effect. 
Very recently, the disorder effect on the Kitaev spin liquid has been argued theoretically \cite{Nasu20,Yamada20,Nasu21}.  It has been shown that the half quantization of $\kappa_{xy}$ is fragile against the site dilution \cite{Nasu20}, which appears to be consistent with the present results.

\begin{figure}[t]
	\begin{center}
		\includegraphics[width=0.85\linewidth]{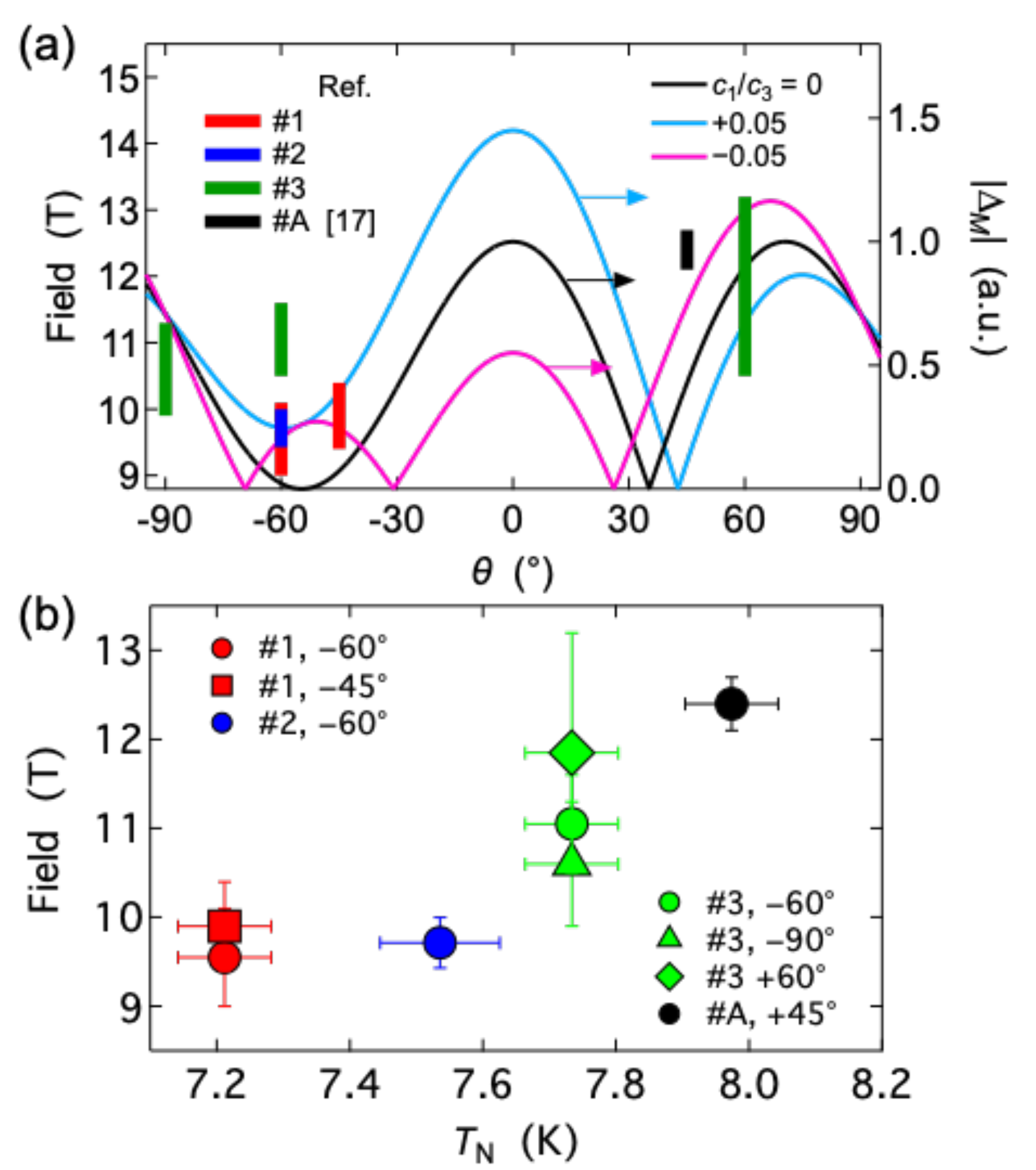}
		\caption{(a) The red, blue, green, and black bars represent the field range (left axis)  where the plateau of HIQTH conductance is observed for samples No.\,1, No.\,2, No.\,3, and A (Ref.\,\cite{Yamashita20}), respectively.  The black solid line represent the Majorana gap $|\Delta_M|$ (right axis) in the absence of non-Kitaev interactions.   Light blue and pink solid lines represent $|\Delta_M|$ in the presence of a non-Kitaev interaction for $c_1/c_3=+0.05$ and $-0.05$, respectively.  (b) The quantization field vs. $T_N$ for samples No.\,1, No.\,2, No.\,3, and A (Ref.\,\cite{Yamashita20}). The symbols represent the mid-points of the plateau field, and error bars represent the onset and offset fields for the HIQTH  plateau. }
	\end{center}
\end{figure}

We here comment on the phonon thermal Hall effect. The observation of the planar thermal Hall effect \cite{Yokoi21,Czajka21} and the sign change of $\kappa_{xy}$ with magnetic field rotation \cite{Yokoi21} cannot be explained by the phonon scenario, but consistent with the Kitaev model. The phonon scenario has been claimed based on the results that the temperature dependence of $\kappa_{xy}$ is similar to that of $\kappa_{xx}$ for {\boldmath $H$}$\parallel c$ below and above $T_N$; both $\kappa_{xx}$ and $\kappa_{xy}$ exhibit a clear upturn at $T_N$ \cite{Lefrancois21}. In sharp contrast, however, as reported in Refs.\,\cite{Kasahara18PRL,Hentrich19},  while $\kappa_{xx}$ exhibits a clear upturn, $\kappa_{xy}$ only changes its sign without showing an upturn, reporting the absence of the trend claimed in Ref.\,\cite{Lefrancois21}. 

We finally discuss the magnetic field range where the HIQTH  plateau is observed.  In Fig.\,5(a), the field range of the plateau at three different angles for sample No.\,3 is shown by the green bars.  The lower boundary of the plateau does not depend much on the angle, but the upper boundary is distinctly higher at $+60^{\circ}$ than at $-90^{\circ}$ or $-60^{\circ}$.  In the presence of non-Kitaev interactions, such as Heisenberg and off-diagonal exchanges, the Chern number is written as, $C_h={\rm sgn}\{c_1(h_x+h_y+h_z)+c_3(h_xh_yh_z)+\cdots\}$, where  $c_1$ and $c_3$ terms represent the contributions of the non-Kitaev and Kitaev terms, respectively \cite{Yokoi21}. In this case, the Majorana gap is written as $\Delta_M/c_3=(c_1/c_3)(h_x+h_y+h_z)+(h_xh_yh_z)$.    In Fig.\,5(a), the angular variation of $|\Delta_M|$ for a pure Kitaev model with $c_1=0$ and in the presence of non-Kitaev interactions with $c_1/c_3=\pm 0.05$ are shown by solid lines. For all $c_1$ values, $|\Delta_M|$ takes a maximum at $\theta\sim+60^\circ$. Although a more accurate theory incorporating the non-Kitaev terms is needed to calculate the detailed angular dependence, the higher upper boundary at  $\theta=+60^{\circ}$  may be related to the larger $|\Delta_M|$.

In Fig.\,5(a), we also plot the magnetic field range of the plateau for samples No.\,1 (red) and No.\,2 (blue), along with the data reported in Ref.\,\cite{Yamashita20}.  The field range of the quantization is sample dependent.   We note that as shown in Fig.\,1(c), the  N\'{e}el temperature is also sample dependent.  To examine the correlation between the quantization field and  N\'{e}el temperature, the quantization field is plotted against $T_N$ in Fig.\,5(b).    Notably, the sample with a higher  N\'{e}el temperature tends to have a higher onset field of the quantization.   As the N\'{e}el temperature is determined by the Kitaev and non-Kitaev interactions, the results suggest that the difference of the plateau fields appears to be associated with the slightly sample-dependent Kitaev and non-Kitaev interactions. While further studies are warranted, the present results provide an important clue to the origin of the sample-dependent quantization field.

In summary, we have measured $\kappa_{xy}$ of various samples grown by the Bridgman method.   The results show that there is a relationship between the HIQTH effect and the thermal conductivity in zero fields.   Samples exhibiting the HIQTH effect have larger zero-field longitudinal thermal conductivity below and above the N\'{e}el temperature than those not exhibiting quantization. In the unquantized samples, $|\kappa_{xy}|/T$ is always smaller than the quantized value and decreases monotonically with decreasing temperature in the FIQD state.   In addition, we find that the onset field of the quantization is nearly angular independent in the same sample but distinctly depends on a slightly sample-dependent N\'{e}el temperature. The samples with a higher N\'{e}el temperature exhibit higher quantization fields.  These results indicate that the quantization phenomenon is significantly influenced by the impurity/defect scatterings and the non-Kitaev interactions.

\begin{acknowledgments}
We thank S. Fujimoto, Y. Motome, and J. Nasu for insightful discussions. This work is supported by Grants-in-Aid for Scientific Research (KAKENHI) (No.\,JP17H01142, No.\,JP18H05227, No.\,JP19H00649, No.\,JP19K03711, No.\,JP21H04443, and No.\,JP21K13881) and on Innovative Areas ``Quantum Liquid Crystals" (No.\,JP19H05824 and No.\,JP19H05825) from the Japan Society for the Promotion of Science, and JST CREST (JPMJCR19T5). 
\end{acknowledgments}

\end{document}